\documentclass[conference]{IEEEtran}
\IEEEoverridecommandlockouts
\usepackage{amsmath,amssymb,amsfonts}
\usepackage{graphicx}
\usepackage{textcomp}
\usepackage{xcolor}
\usepackage{svg}
\def\BibTeX{{\rm B\kern-.05em{\sc i\kern-.025em b}\kern-.08em
    T\kern-.1667em\lower.7ex\hbox{E}\kern-.125emX}}
\usepackage{cite}
\usepackage{comment}

\usepackage{nohyperref}

\definecolor{matplotlib0}{HTML}{1f77b4}
\definecolor{matplotlib1}{HTML}{d62728}
\definecolor{matplotlib2}{HTML}{2ca02c}
\definecolor{matplotlib3}{HTML}{ff7f0e}
\definecolor{matplotlib4}{HTML}{9467bd}
\definecolor{matplotlib5}{HTML}{8c564b}
\definecolor{matplotlib6}{HTML}{e377c2}
\definecolor{matplotlib7}{HTML}{7f7f7f}
\definecolor{matplotlib8}{HTML}{bcbd22}
\definecolor{matplotlib9}{HTML}{17becf}

\usepackage{mathtools} % \abs, ...

\usepackage{booktabs}
\usepackage{multirow}
\usepackage{colortbl}
\usepackage{tablefootnote}
\usepackage{threeparttable}

\usepackage[acronym, style=super, nonumberlist]{glossaries}

\usepackage{pgfplots}
\usepgfplotslibrary{fillbetween}
\usepgfplotslibrary{colormaps}
\pgfplotsset{compat=1.16}

\pgfplotscreateplotcyclelist{matplotlib}{
  {matplotlib0},
  {matplotlib1},
  {matplotlib2},
  {matplotlib3},
  {matplotlib4},
  {matplotlib5},
  {matplotlib6},
  {matplotlib7},
  {matplotlib8},
  {matplotlib9}
}

\pgfplotsset{every axis/.append style={
    cycle list name=matplotlib,
}}

\usepackage{listings}

\definecolor{code_default}{HTML}{000000}
\definecolor{code_keyword}{HTML}{AC4142}
\definecolor{code_identifier}{HTML}{D28445}

\lstdefinelanguage{RISCV}{
  sensitive=false,
  morecomment=[l]{//},
  alsoletter={.},
  morekeywords=[1]{
    lp.setup, mv, lw, p.lw, sw, p.sw, pv.sdotsp.b, pv.shuffle2.b, p.subNR, p.addNR
  },
  morekeywords=[2]{
    zero, ra, sp, gp, tp, t0, t1, t2, t3, t4, t5, t6, s0, s1, a0, a1, a2, a3, a4, a5, a6, a7, a8, a9, a10, a11,
  },
  morestring=[b]",
  morestring=[b]',
}[strings, comments, keywords]

\lstdefinestyle{RISCV_STYLE}{
  language=RISCV,
  numbers=none,
  basicstyle=\scriptsize\ttfamily\color{code_default},
  keywordstyle=[1]\color{matplotlib0},
  keywordstyle=[2]\color{matplotlib1},
  float,
  captionpos=b,
  belowskip=-0.5cm
}

\usepackage{algorithm}
\usepackage{algpseudocode}
\usepackage{float}
\newfloat{algorithm}{t}{top}

\newacronym{simd}{SIMD}{Single Instruction, Multiple Data}
\newacronym{elu}{ELU}{Exponential Linear Unit}
\newacronym{relu}{ReLU}{Rectified Linear Unit}
\newacronym{rpr}{RPR}{Random Partition Relaxation}
\newacronym{mac}{MAC}{Multiply Accumulate}
\newacronym{dma}{DMA}{Direct Memory Access}
\newacronym{bmi}{BMI}{Brain--Machine Interface}
\newacronym{bci}{BCI}{Brain--Computer Interface}
\newacronym{smr}{SMR}{Sensory Motor Rythms}
\newacronym{eeg}{EEG}{Electroencephalography}
\newacronym{svm}{SVM}{Support Vector Machine}
\newacronym{svd}{SVD}{Singular Value Decomposition}
\newacronym{evd}{EVD}{Eigendecomposition}
\newacronym{iir}{IIR}{Infinite Impulse Response}
\newacronym{fir}{FIR}{Finite Impulse Response}
\newacronym{fc}{FC}{Fabric Controller}
\newacronym{nn}{NN}{Neural Network}
\newacronym{mrc}{MRC}{Multiscale Riemannian Classifier}
\newacronym{flop}{FLOP}{Floating Point Operation}
\newacronym{sos}{SOS}{Second-Order Section}
\newacronym{ipc}{IPC}{Instructions per Cycle}
\newacronym{tcdm}{TCDM}{Tightly Coupled Data Memory}
\newacronym{fpu}{FPU}{Floating Point Unit}
\newacronym{fma}{FMA}{Fused Multiply Add}
\newacronym{alu}{ALU}{Arithmetic Logic Unit}
\newacronym{dsp}{DSP}{Digital Signal Processing}
\newacronym{gpu}{GPU}{Graphics Processing Unit}
\newacronym{soc}{SoC}{System-on-Chip}
\newacronym{mi}{MI}{Motor-Imagery}
\newacronym{csp}{CSP}{Commmon Spatial Patterns}
\newacronym{fbcsp}{FBCSP}{Filter-Bank \acrlong{csp}}
\newacronym{pulp}{PULP}{parallel ultra-low power}
\newacronym{soa}{SoA}{state-of-the-art}
\newacronym{bn}{BN}{Batch Normalization}
\newacronym{isa}{ISA}{Instruction Set Architecture}
\newacronym{ecg}{ECG}{Electrocardiogram}
\newacronym{mcu}{MCU}{microcontroller}
\newacronym{rnn}{RNN}{recurrent neural network}
\newacronym{cnn}{CNN}{convolutional neural network}
\newacronym{tcn}{TCN}{temporal convolutional network}
\newacronym{emu}{EMU}{epilepsy monitoring unit}

\renewcommand{\baselinestretch}{0.95}

\newcommand{\luca}[1]{{\color{black}#1}}

\makeatletter
\def\ps@IEEEtitlepagestyle{%
  \def\@oddfoot{\mycopyrightnotice}%
  \def\@oddhead{\hbox{}\@IEEEheaderstyle\leftmark\hfil\thepage}\relax
  \def\@evenhead{\@IEEEheaderstyle\thepage\hfil\leftmark\hbox{}}\relax
  \def\@evenfoot{}%
}
\def\mycopyrightnotice{%
  \begin{minipage}{\textwidth}
  \centering \scriptsize
  \copyright 2021 IEEE.  Personal use of this material is permitted.  Permission from IEEE must be obtained for all other uses, in any current or future media, including reprinting/republishing this material for advertising or promotional purposes, creating new collective works, for resale or redistribution to servers or lists, or reuse of any copyrighted component of this work in other works.
  \end{minipage}
}
\makeatother

\begin{document}

\title{
Towards Long-term Non-invasive Monitoring for Epilepsy via Wearable EEG Devices
}

 \author{\IEEEauthorblockN{
    Thorir Mar Ingolfsson\IEEEauthorrefmark{1},
    Andrea Cossettini\IEEEauthorrefmark{1},
    Xiaying Wang\IEEEauthorrefmark{1}, 
    Enrico Tabanelli\IEEEauthorrefmark{2},
    Giuseppe Tagliavini\IEEEauthorrefmark{2},\\
    Philippe Ryvlin\IEEEauthorrefmark{4},
    Luca Benini\IEEEauthorrefmark{1}\IEEEauthorrefmark{2},
    Simone Benatti\IEEEauthorrefmark{2}\IEEEauthorrefmark{3}}
     
    \vspace{0.2cm}

    \IEEEauthorblockA{\IEEEauthorrefmark{1}Integrated Systems Laboratory, ETH Z{\"u}rich, Z{\"u}rich, Switzerland}
    \IEEEauthorblockA{\IEEEauthorrefmark{2}DEI, University of Bologna, Bologna, Italy}
    \IEEEauthorblockA{\IEEEauthorrefmark{3}University of Modena and Reggio Emilia, Modena, Italy}
    \IEEEauthorblockA{\IEEEauthorrefmark{4}Lausanne  University  Hospital (CHUV), Switzerland}
    
    \thanks{Corresponding email: \{thoriri\}@iis.ee.ethz.ch}
    
    \vspace{-0.5cm}
    }

\maketitle

\begin{abstract}
We present the implementation of seizure detection algorithms based on a minimal number of EEG channels on a parallel ultra-low-power embedded platform. The analyses are based on the CHB-MIT dataset, and include explorations of different classification approaches (Support Vector Machines, Random Forest, Extra Trees, AdaBoost) and different pre/post-processing techniques to maximize sensitivity while guaranteeing no false alarms. We analyze global and subject-specific approaches, considering all 23-electrodes or only 4 temporal channels. For 8\,s window size and subject-specific approach, we report zero false positives and 100\% sensitivity. These algorithms are parallelized and optimized for a parallel ultra-low power (PULP) platform, enabling 300h of continuous monitoring on a 300 mAh battery, in a wearable form factor and power budget. These results pave the way for the implementation of affordable, wearable, long-term epilepsy monitoring solutions with low false-positive rates and high sensitivity, meeting both patient and caregiver requirements.
\end{abstract}
\begin{IEEEkeywords}
healthcare, time series classification, smart edge computing, machine learning, deep learning.
\end{IEEEkeywords}

\section{Introduction}
\label{sec:introduction}
Epilepsy is a brain disease that affects more than 50 million people worldwide~\cite{world2019epilepsy}.
Conventional treatments are mostly pharmacological, but they can require surgery or invasive neurostimulation in the case of drug-resistant subjects.
To personalize patient treatments, continuous monitoring of the brain activity is required, and it can be performed in \glspl{emu}, where patients are monitored by a video surveillance system synchronized with 32-64 channels EEG caps. \gls{emu} monitoring helps clinicians to refine the diagnosis of the seizures and to monitor the effects of therapies in the short term (i.e., 1--2 weeks). However, continuous monitoring of the patients outside the \gls{emu} is paramount to control and refine therapies in the medium period. 

The golden diagnostic standard is represented by \gls{eeg} systems, which unfortunately are cumbersome and can make patients uncomfortable because of perceived stigmatization. Thus, both patients and caregivers would benefit from the availability of wearable long-term \gls{eeg} monitoring devices,
which can also detect sub-clinical seizures with high levels of sensitivity and low false detection rate~\cite{bruno2020seizure}.

This solution poses several challenges. Firstly, traditional \gls{eeg} caps have 16--32 channels, providing full coverage of the scalp. Hence, it must be understood if it is possible to effectively detect seizures with a reduced electrode set, to be placed into minimally intrusive devices, such as glasses or behind-the-ear systems~\cite{asif2020epileptic,pham2020wake,sopic2018glass,guermandi2018wearable}.
The second challenge is linked to the performance of the system, especially for what concerns specificity. False alarms represent one of the major causes of stress in seizure detection frameworks, because they raise the level of anxiety of the subject, which has been identified to be a potential trigger for seizures~\cite{MCKEE201721stress}. Recent works~\cite{burrello2019laelaps} provide epilepsy detection algorithms with very low false-alarms, but they are based on computationally demanding algorithms and are used in implantable systems for pre-surgical monitoring.

To address the first challenge, investigations about applying machine learning (ML) algorithms on reduced EEG montages are needed. In particular, the focus should be put on meeting the limited computation resources of embedded platforms.
As regards the second challenge, parallel and ultra-low-power (PULP) embedded platforms represent a key enabling technology for the successful deployment of such algorithms into wearable long-term monitoring solutions~\cite{pullini2019mr}, since they overcome the limits of commercial devices in terms of power consumption, while providing high computational capabilities. 

Within these considerations, this work presents a seizure detection framework designed for wearable unobtrusive systems. Specifically, we propose the following contributions:
\begin{itemize}
    \item A comparison over several state-of-the-art (SoA) detection algorithms (namely, Support Vector Machines, Random Forest, Extra Trees, and AdaBoost), on a benchmark EEG dataset, showing how performance changes between full-coverage (23 channels) and minimal (4 channels over the temporal lobe, \emph{temporal channels} as a shorthand in the remaining text) setup.
    \item An investigation on the impact of different pre- and post-processing approaches to maximize sensitivity and minimize the false-positive (FP) rate. This includes analyzing the performance improvements when relying on subject-specific training instead of global models. 
    \item The \luca{implementation and the performance tuning of the above framework on real PULP chip target}, namely Mr. Wolf~\cite{pullini2019mr}, designed in 40nm FDSOI technology and widely used in ultra-low-power wearable nodes for biomedical applications~\cite{kartsch2019biowolf}. 
\end{itemize}
Experimental results demonstrate that a high-accuracy low-FP rate framework for continuous epilepsy monitoring can be designed in a wearable \luca{long-lifetime system} with a low electrode count. 

\section{Materials \& Methods}
 
\begin{figure}[t]
  \centering
  \includegraphics[width=0.9\linewidth, clip, trim={0 0.15cm 0 0}]{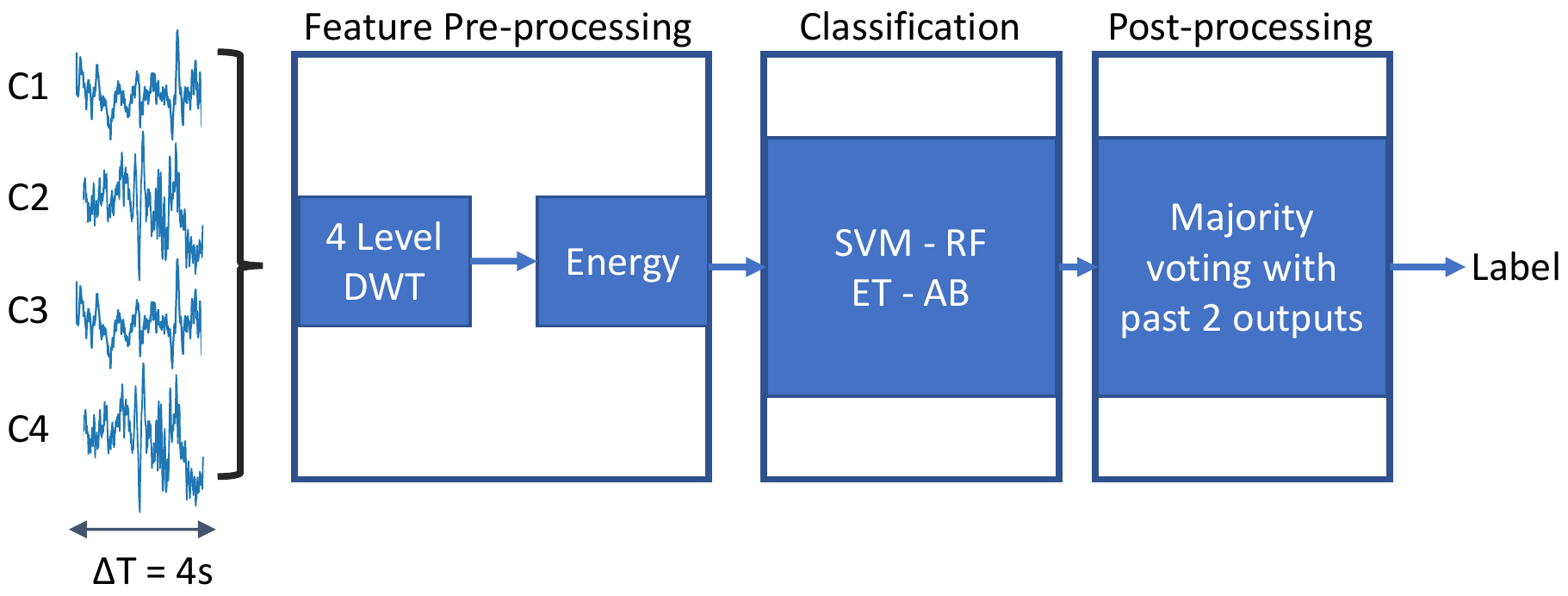}
  \vspace{-0.2cm}
  \caption{Schematic representation of the workflow. Input data (all 23 channels or 4 temporal channels) are pre-processed by means of DWT. Classification is performed with SVM, RF, ET, or AB algorithms. The output of the classifier is post-processed with a majority voting.}
  \label{fig:workflow_sketch}
  \vspace{-0.2cm}
\end{figure}

Our investigations are based on the CHB-MIT dataset from the Children’s Hospital of Boston and MIT~\cite{shoeb2009application, goldberger2000physiobank}.
Fig. \ref{fig:workflow_sketch} shows a sketch of the overall workflow of our investigations, whose details are elaborated in the following sections.
%FFFFFFFFFFFFFFFFFFFFFFFFFFFFFFFFFFFFFFFFFFFFFFFFFFFFFFFFFFFFFFFFFFFFF
% FIGURE
%FFFFFFFFFFFFFFFFFFFFFFFFFFFFFFFFFFFFFFFFFFFFFFFFFFFFFFFFFFFFFFFFFFFFF
\subsection{Pre-processing and feature extraction}
Discrete wavelet transform (DWT) is a  signal decomposition technique for time-frequency analyses~\cite{mallat1989theory}, widely used in preprocessing stages of ML algorithms. In contrast with other signal transforms (e.g., Fourier transforms),  DWT captures both frequency and temporal features of a given signal by applying a cascade series of filters, shaped as \textit{Mother Wavelets}. Input signals are passed iteratively through low-pass and a high-pass filters, resulting in a series of \textit{Approximation} and \textit{Detail} coefficients, respectively. Both filters can be implemented as convolutions between the signal and pre-computed kernels, followed by downsampling by two. Approximation coefficients of a given decomposition level represent the input of the next level; following this pattern, each level halves the time resolution but doubles the frequency resolution.

In the context of bio-signal analysis, DWT has been used in state-of-the-art research works to extract features from physiological data since they provide a good trade-off between performance and signal-to-noise ratio~\cite{benatti2014towards}.
We use DWT with three different window lengths: 2\,s, 4\,s, or 8\,s. Our multi-core platform (Sec.~\ref{sect:mrwolf}) allows to process these different window sizes in a fast, energy efficient, and accurate manner, performing the 4-level DWT on two cores for each channel.

\subsection{Classification}
We target the deployment of memory efficient and low computational algorithms to enable lightweight epileptic seizure detection at the edge.
Thus, we estimate the performance of 4 supervised learning methods: a Support Vector Machine (SVM), Random Forest (RF), Extra Trees (ET), and AdaBoost (AB) classifiers. 
Supervised learning relies on two phases: the training, where a model is tuned by using labeled data (we use ictal and interictal EEG samples to create the models), and the inference, where, exploiting the trained model, the algorithm determines which class a sample belongs to. 
In the following, we give a brief overview of the algorithms mentioned above.\\
\emph{1) SVM} is a supervised ML method that offers high generalization performance. 
By learning a set of hyperplanes in high-dimensional space, SVM separates the feature space and classifies instances. 
To make data linearly separable, we map them into high-dimensional feature spaces deploying the Radial Basis Function (RBF).
The RBF kernel consists of a real-valued function relying on the distance between the input and a fixed point. 
The distance metric is usually the Euclidean Distance. During inference the SVM classifies instances based on which side of the hyperplane they lie.
Referring to the UCI epileptic dataset~\cite{rg2001bonn}, a 98.18\% classification accuracy was achieved by \cite{raut2021comparative} when applying SVM. 
\\
\emph{2) RF} is an easy-to-use widely-adopted ML algorithm that relies on an ensemble of low-correlation tree-structured classifiers~\cite{breiman2001rf}.
By randomly picking a feature space subset at each tree splitting node, the training algorithm  searches for the most performing thresholds to split training instances until reaching a leaf node.
By aggregating trees’ votes, the model inference returns the input prediction picking up the class with the largest number of votes.
A 99.78\% accuracy was achieved by \cite{raut2021comparative} on the UCI dataset.
\\
\emph{3) ET} is a tree-based ensemble classification method that leverages stronger randomization to achieve a favorable bias-variance trade-off~\cite{guerts2006et}.
ET picks a random feature space subset at each node while it randomly generates a threshold for each candidate feature, selecting the optimal one as the node splitting rule.
The additional randomness reduces the complexity of the induction process, increases the training speed, and weakens correlation. ET follows the same approach as RF for model inference.
An ET-based epileptic signal classification algorithm was proposed by \cite{dinghan2020et}, reaching up to 89.49\% seizure detection accuracy on the CHB-MIT dataset.
\\
\emph{4) AB} is a statistical classification meta-algorithm that combines a set of weak learners through a weighted vote to reduce bias and variance in supervised learning tasks~\cite{freund1997decision}.
The boosting technique enables learning an ensemble algorithm with higher performance than single learners and is less prone to overfitting. During inference, AB sums up the weighted vote of the weak learners to classify between instances.
An automatic seizure detection algorithm based on AB was shown to reach 99.08\% accuracy on the UCI epileptic dataset \cite{hassan2020epilepsy}.

As in~\cite{shah2020validation}, during the training phase we heavily penalize the false alarm rate by assigning different weights to \textit{seizures} and \textit{non-seizures}. We define the class weights as the inverse of the frequency of the occurrence of the two classes. In this configuration, the weight ratios are "balanced". We perform classifications considering the following scenarios:
\begin{itemize}
    \item training/testing on all patients, using all 23 channels
    \item training/testing on all patients, with 4 temporal channels
    \item subject-specific training/testing, using all 23 channels
    \item subject-specific training/testing, with 4 temporal channels
\end{itemize}

\subsection{Post-processing}
Once occurring, epileptic seizures are expected to last multiple seconds. As in~\cite{pale2021systematic}, we apply moving-averages to the labels with a window of 4s, performing majority voting, thus effectively smoothing out the predictions. 
\luca{The low-pass filtering effect of the moving average gives a great benefit in terms FP reduction, since it basically eliminates the fluctuations of the classifier output. In the following, we consider moving averages over 3 successive classifications.}
 
\subsection{Embedded platform}
\label{sect:mrwolf}

\begin{figure}[h!]
  \centering
  \vspace{-0.3cm}
  \includegraphics[width=0.58\linewidth, clip, trim={0 0 0 0}]{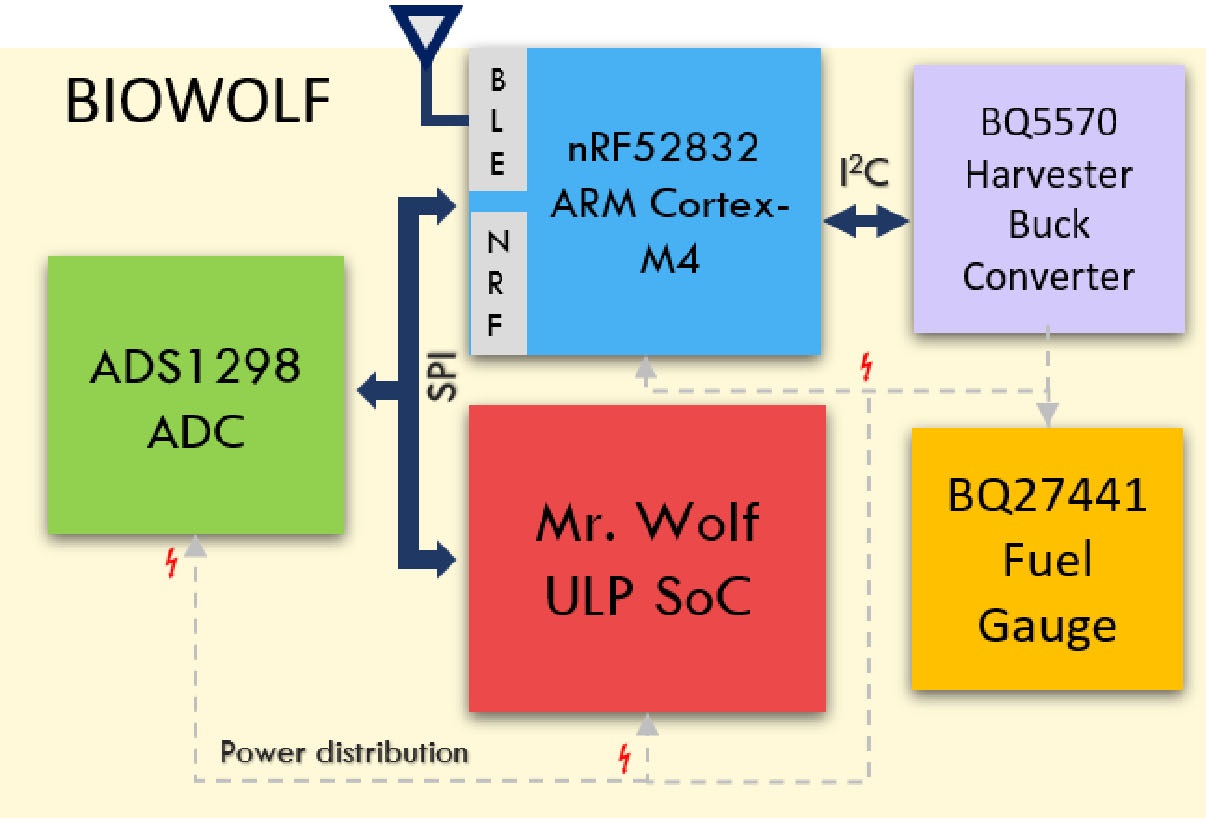}
  \includegraphics[width=0.29\linewidth, clip, trim={0 0 0 0}]{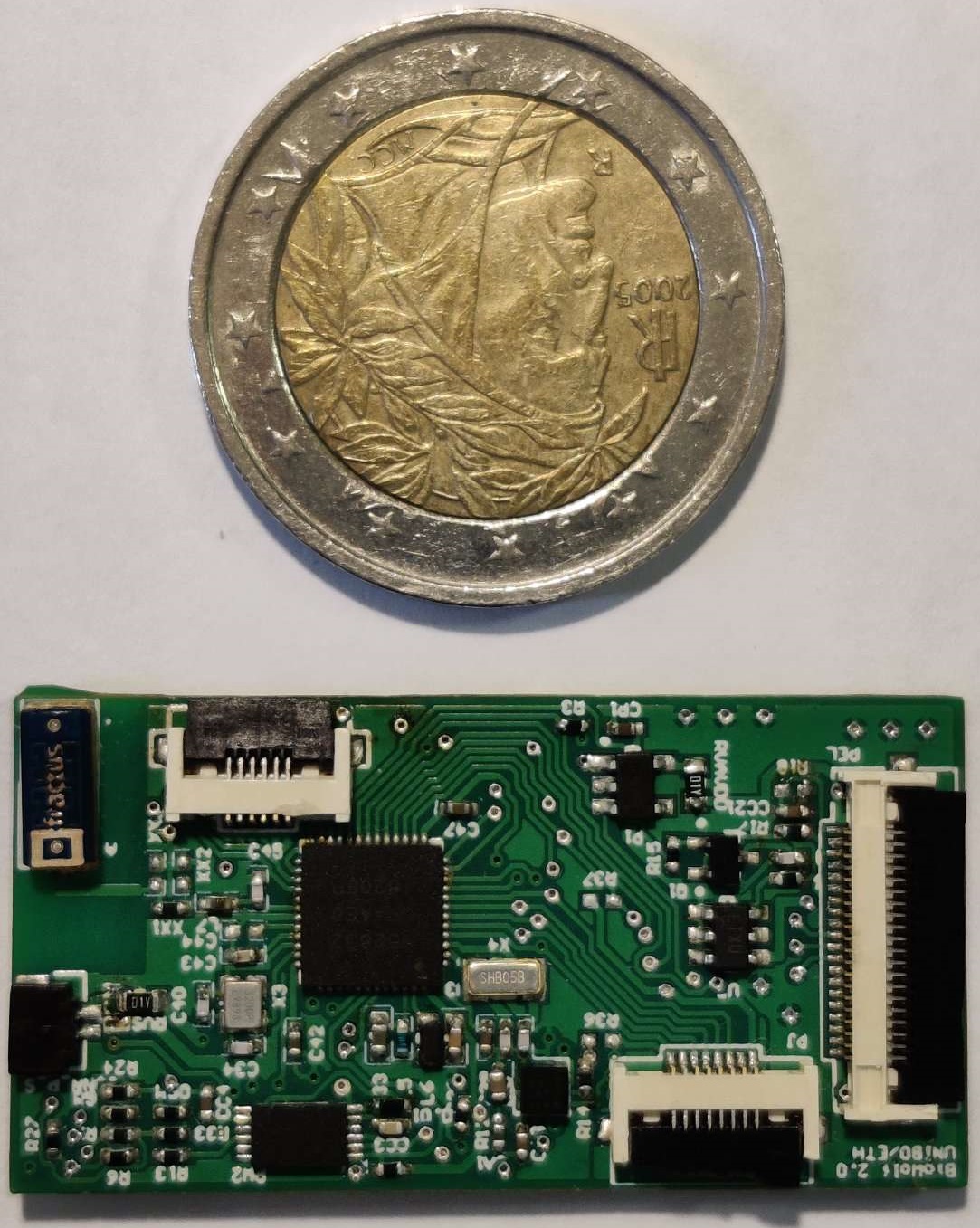}
  \vspace{-0.2cm}
  \caption{Left: main building blocks of the BioWolf~\cite{kartsch2019biowolf} embedded platform. Right: \luca{wearable form factor of BioWolf, compared to a 2 euros coin.}}
  \label{fig:MrWolf}
  \vspace{-0.2cm}
\end{figure}

The seizure detection framework is implemented on the \luca{BioWolf wearable ExG device~\cite{kartsch2019biowolf}} (Fig. \ref{fig:MrWolf}). BioWolf includes an 8-channel analog front-end (AFE) for biopotential acquisition, the PULP Mr. Wolf multicore processor, and a wireless Bluetooth low-energy (BLE) link enabled by a Nordic nRF8232 chip. By virtue of the multicore digital platform, BioWolf outperforms \luca{by at least one order of magnitude} the performance achievable with conventional single-core low-power processors, such as ARM CORTEX M4 or MSP 430 based devices, \luca{with a comparable power budget}.  The BioWolf system is based on a fully programmable processor, which combines versatility of a programmable system, where algorithms can be tuned and improved, with the efficiency of a computational 8-core cluster with 2 shared floating point units. It should also be noted that a \luca{hardwired application specific integrated circuit (IC) solution is more efficient (at the price of lack of flexibility)} only in applications where the power envelope is very critical (e.g., implantable devices), and typically can not exceed 1--2\,mW total power.
%FFFFFFFFFFFFFFFFFFFFFFFFFFFFFFFFFFFFFFFFFFFFFFFFFFFFFFFFFFFFFFFFFFFFF
%FFFFFFFFFFFFFFFFFFFFFFFFFFFFFFFFFFFFFFFFFFFFFFFFFFFFFFFFFFFFFFFFFFFFF
 \subsection{Power Measurements}
We perform the measurements on Mr. Wolf considering cluster domain with the processors running at 100 MHz (frequency of maximal energy efficiency~\cite{pullini2019mr}). The measurements are performed using a Keysight N6705C power analyzer, with 0.2048\,ms sampling interval.

\section{Results \& Discussion}
\label{sec:results}
Table \ref{tab:results:summary} summarizes the \luca{accuracy metrics} reported for the four models considered in this paper, applied on the CHB-MIT data. We compare a global and a subject-specific approach to test the models.
The reported numbers correspond to a balanced weighting ratio. In the global approach, SVM and AB have too high\,FPs/h. RF and ET have fewer\,FPs/h, but lack the sensitivity to be considered reliable approaches. These conclusions apply both when considering all 23 channels (10-20 EEG standard) and when considering only the 4 temporal ones (F7-T7, T7-P7, F8-T8, T8-P8). 

\begin{figure}[t]
  \centering
  \vspace{-0.0cm}
  \includegraphics[width=\columnwidth]{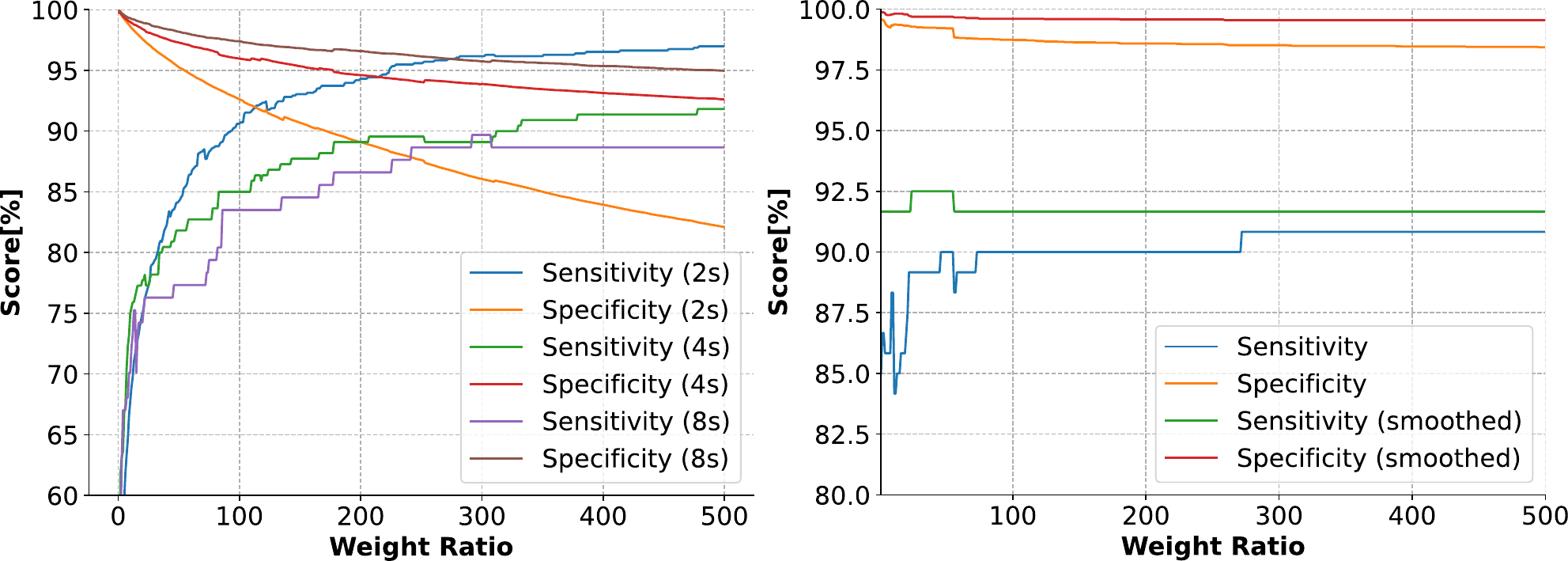}
  %\includesvg[width = \columnwidth]{Figures/global_adaboost_vs_temporal_2s_4s_8s}
  \vspace{-0.2cm}
  \caption{Sensitivity/Specificity plot for subject-specific AdaBoost. Left: global model approach, demonstrating the dependencies on the weighting ratio and on the choice of the window length (2\,s, 4\,s, or 8\,s). Increasing the weight ratio results in greater sensitivity, at the cost of a reduced specificity. Longer windows improve specificity, at the cost of a reduced sensitivity. With weight ratio = 100, the obtained FP rate is 133\,FP/h, 36\,FP/h, and 11\,FP/h, for the 2\,s, 4\,s, and 8\,s windows, respectively. Right: subject-specific approach, demonstrating the benefit of post-processing of output. For both plots the electrodes considered are 4 temporal channels.}
  \label{fig:AdaBoost_smoothing}
\end{figure}

On the contrary, in subject-specific training, specificity and sensitivity increase for all models, providing at the same time a reduced FP rate.
Fig. \ref{fig:AdaBoost_smoothing} shows that the performance of AB increases (both in terms of specificity and sensitivity) at increasing temporal window lengths in the feature pre-processing; the other algorithms provide a similar behavior (Tab. \ref{tab:results:summary}), where for the subject-specific training we approach 0\,FP/h, while retaining high sensitivity values.
Fig. \ref{fig:AdaBoost_smoothing} also shows the effect of post-processing (smoothing) the classification outputs: both specificity and sensitivity are increased. We notice that, at least for data analyzed in this experiments, reducing the number of electrodes does not affect significantly classification performance. As a result, the deployment of a minimally invasive setup, which can be integrated in smart-glasses or other wearable devices with low stigmatization, seems to be feasible. 

\begin{table*}[ht!]
\renewcommand{\arraystretch}{1.1}
  \centering
  \caption{Comparison between algorithms}\label{tab:results:summary}
  \vspace{-.2cm}
  {
\begin{tabular}{@{}llrrrrrrrrrrrrrrrr@{}}
\toprule
&& \multicolumn{4}{c}{All Channels} & \multicolumn{12}{c}{Temporal Channels} \\
\cmidrule(lr){3-6} \cmidrule(l){7-18}
&& \multicolumn{4}{c}{2s window, w/o smooth.} & \multicolumn{4}{c}{2s window, w/o smooth.} & \multicolumn{4}{c}{8s window, w/o smooth.} & \multicolumn{4}{c}{8s window, w smooth.}\\
\cmidrule(lr){3-6} \cmidrule(lr){7-10} \cmidrule(lr){11-14} \cmidrule(l){15-18}
 & & \multicolumn{1}{r}{SVM}   & \multicolumn{1}{r}{RF}    & \multicolumn{1}{r}{ET}    & \multicolumn{1}{r}{AB} & \multicolumn{1}{r}{SVM}   & \multicolumn{1}{r}{RF}    & \multicolumn{1}{r}{ET}    & \multicolumn{1}{r}{AB} & \multicolumn{1}{r}{SVM}   & \multicolumn{1}{r}{RF}    & \multicolumn{1}{r}{ET}    & \multicolumn{1}{r}{AB} & \multicolumn{1}{r}{SVM}   & \multicolumn{1}{r}{RF}    & \multicolumn{1}{r}{ET}    & \multicolumn{1}{r}{AB} \\ 
\midrule
\multirow{3}{*}{\rotatebox[origin=c]{90}{\parbox[c]{1cm}{\centering \textbf{Global}}}}&\hspace{1mm}Specificity [\%]         & 95.9 & 99.9 & 99.9 & 94.0 & 96.3 & 99.9 & 99.9 & 98.5   & 96.3 & 99.9 &  99.9 & 98.1  & 97.1 & 99.9 &  99.9 & \textbf{99.1}\\
&\hspace{1mm}Sensitivity [\%]        & 72.7 & 39.0 & 34.8 & 84.6 & 75.1 & 41.2 & 35.2 & 84.3 &75.1 & 48.5 & 42.3 & 77.3 & 81.1 & 49.5 & 47.3 & \textbf{85.3}  \\
&\hspace{1mm}FP/h                 & 73.5 & 1.8   & 1.6   & 107.8  &65.9  & 2.2   & 1.6   & 27.2    & 16.5 & 0.4 & 0.09 & 8.3 & 13.1 & 0.4 & 0.09 & \textbf{3.6} \\ 
\midrule
\multirow{3}{*}{\rotatebox[origin=c]{90}{\parbox[c]{1cm}{\centering \textbf{Subject-Specific}}}}&\hspace{1mm}Specificity [\%]          & 97.9   & 99.9   & 99.9   & 99.4 & 99.1   & 99.8   & 99.8   & 99.4   & 98.9 & 99.9 & 99.8 & 99.9 & \textbf{99.4} & \textbf{100} & \textbf{100} & \textbf{100} \\ 
&\hspace{1mm}Sensitivity [\%]         & 84.2   & 63.6   & 65.1   & 87.8 & 88.9   & 86.7   & 89.2   & 88.9  & 100 & 100 & 100 & 100 & \textbf{100} & \textbf{100} & \textbf{100} & \textbf{100}\\
&\hspace{1mm}FP/h                 & 37.8    & 2.2     & 1.8     & 10.3  & 17.1    & 3.8     & 2.5     & 10.2  & 5.1 & 0.5 & 1.0 & 0.4 & \textbf{2.7} & \textbf{0} & \textbf{0} & \textbf{0}\\
\bottomrule
\end{tabular}
  }
  \vspace{-0.5cm}
\end{table*}

%FFFFFFFFFFFFFFFFFFFFFFFFFFFFFFFFFFFFFFFFFFFFFFFFFFFFFFFFFFFFFFFFFFFFF
%FFFFFFFFFFFFFFFFFFFFFFFFFFFFFFFFFFFFFFFFFFFFFFFFFFFFFFFFFFFFFFFFFFFFF

Regarding the embedded deployment of the classification algorithms, the BioWolf platform benefits of the multicore computing cluster of the PULP chip.    
Tab.~\ref{tab:results:summary:wolf} summarizes 
energy measurements 
and parallel speedup concerning sequential execution for the four algorithms evaluated on Mr. Wolf. RF, ET, and AB achieve near-ideal speedups between 6.89-7.21$\times$, only limited by architetural factors, such as contentions due to the concurrent accesses to tightly coupled data memory (TCDM) banks and instruction misses on the instruction cache, which does not scale linearly with the cores.
Instead, SVM consists of a workload parallelizable up to around 97\%, leading to an ideally achievable speedup of 6.53$\times$. Therefore, it reaches a lower speedup of 5.84$\times$, also bounded by the aforementioned architectural limits. RF, ET, and AB then show much better energy usage than the SVM, requiring only around 1\,\textmu J per inference. This stems from the much less resource-intensive calculations needed than for decision-tree-based inference. In comparison, assuming an SoA commercial AFE for biosignal acquisition such as T.I. ADS1298, which requires 0.75\,mW per channel, and considering a battery of 300\,mAh, our approaches implemented on Mr. Wolf would last for roughly 300 hours of data acquisition and classification at 4\,s intervals. Comparing to~\cite{zanetti2020robust}, which implemented a seizure classification method on an STM32L476 ARM Cortex-M4 microcontroller, we see a 7.34$\times$ improvement in battery life.

%FFFFFFFFFFFFFFFFFFFFFFFFFFFFFFFFFFFFFFFFFFFFFFFFFFFFFFFFFFFFFFFFFFFFF
\begin{table}[t]
\renewcommand{\arraystretch}{1}
  \centering
  \caption{Comparison Between AdaBoost, Random Forest, Extra Trees and Support Vector Machine on Mr Wolf.}\label{tab:results:summary:wolf}
  \vspace{-.2cm}
  {
    \footnotesize
    \begin{tabular}{@{}lrrrr@{}}
      \toprule
      Classifier & SVM & RF & ET & AB \\
      \midrule
      Time/inference [ms]    &0.20 &\textbf{0.052} & \textbf{0.052} & 0.057 \\
      Power [mW] & 25.41 &\textbf{21.89} & \textbf{21.89} & 22.44 \\
      Energy/inference [{\textmu}J]   &5.08 &\textbf{1.14} & \textbf{1.14} & 1.28\\
      \luca{Parallel speedup} & \luca{5.84$\times$} & \luca{\textbf{7.21$\times$}} & \luca{\textbf{7.21$\times$}}& \luca{6.89$\times$} \\
      \bottomrule
    \end{tabular}
  }
\end{table}

%FFFFFFFFFFFFFFFFFFFFFFFFFFFFFFFFFFFFFFFFFFFFFFFFFFFFFFFFFFFFFFFFFFFFF

The implementations of RF, EF, and AB on a multi-core edge platform, such as Mr. Wolf, require a power envelope of only around 22\,mW with sub-100\,\textmu s processing time, thus with lower energy requirements than the AFE and BLE of the complete platform that enables multi-day functionality~\cite{kartsch2019biowolf}.
These results pave the way to the desing of unobtrusive EEG wearable devices that can help patient and clinicians in long term monitoring of epileptic seizures in out-of-the-hospital environments.

\vspace{-0.1cm}
\section{Conclusion}\label{ch:conclusion}
\vspace{-0.1cm}
This work presents a systematic analysis on the feasibility of long-term epilepsy monitoring with minimal EEG setups. We used DWT for signal pre-processing, in combination with four different classification algorithms (SVM, Random Forest, Extra Trees, AdaBoost), to assess the impact of limiting the number of EEG channels to only 4 covering the temporal lobe, enabling a discrete and user comfortable setup. Analyses are done on the CHB-MIT dataset, and we demonstrate that with proper weighting, selection of the observation window size, and smoothing of predictions, 100\% sensitivity and zero false-positives can be achieved, when the algorithms are trained and tested on the same subjects. We further deploy the algorithms on a PULP platform, resulting in minimal energy requirements ($\approx$ 1\,{\textmu}J per inference), outperforming competing commercial devices. These results demonstrate that a PULP system is indeed one of the best candidates for future wearable epilepsy monitoring systems based on minimal EEG setups to avoid stigmatization. Future work will focus also on expanding the proposed techniques to include sensor fusion from other data sources, such as electrocardiograms~\cite{thoriraicas} or inertial sensors, to be realized in a wearable setting of a body area network and on testing the whole system in an ambulatory  and domestic environments.
\vspace{-0.1cm}
\section*{Acknowledgment}
\vspace{-0.1cm}
This project was supported by the Swiss National Science Foundation (Project PEDESITE) under grant agreement 193813.

%%%%%%%%%%%%%%%%%%%%%%%%%%%%%%%%%%%%%%%%%%%%%%%%%%%%%%%%%%%%%%%%%%%%%%%%%%%
%%%%%%%%%%%%%%%%%%%%%%%%%%%%%%%%%%%%%%%%%%%%%%%%%%%%%%%%%%%%%%%%%%%%%%%%%%%
\bibliographystyle{IEEEtran}
\bibliography{references_eeg.bib}

\end{document}